\documentclass[10pt, a4paper]{amsart}

\usepackage{amsaddr}
\usepackage[english]{babel}
\usepackage[utf8x]{inputenc}
\usepackage{amsfonts}
\usepackage{mathtools}
\usepackage{amsthm}
\usepackage{pict2e}
\usepackage{xcolor}
\theoremstyle{plain}
\newtheorem{thm}{Theorem}[section]

\theoremstyle{definition}

\theoremstyle{remark}

\numberwithin{equation}{section}

\title{Towards a full general relativistic approach to galaxies}

\author{Davide Astesiano, Sergio L. Cacciatori \& Federico Re}
\address{Department of Science and High Technology, Universit\`{a}  dell'Insubria,\\ Via Valleggio 11, 22100, Como, Italy\\ \&
INFN, sezione di Milano, Via Celoria 16, 20133, Milano, Italy}

\begin{document}

\begin{abstract}
We analyze the dynamics of a single spiral galaxy from a general relativistic viewpoint. We employ the known family of stationary axially-symmetric solutions to Einstein gravity coupled with dust in order to model the halo external to the bulge.
In particular, we generalize the known results of Balasin and Grumiller, relaxing the condition of co-rotation, thus including non co-rotating dust. This further highlights the discrepancy between Newtonian theory of gravity and general relativity at low velocities
and energy densities. We investigate the role of dragging in simulating dark matter effects. In particular, we show that non co-rotance further reduce the amount of energy density required to explain the rotation curves for spiral galaxies.  
\end{abstract}


\maketitle
\flushbottom

\section{Introduction}
Applying Newton's theory to describe the dynamics of galaxies leads to many differences with observational data.
One of the main problems resides in the non-Keplerian velocity profile: far from their center, spiral galaxies show an almost flat rotation curve. In the recent years there have been several attempts to conciliate these facts with theory. On one hand,
the efforts to justify this discrepancy include the modification of the gravitational theory itself. In this direction we have the so-called MOND theory \cite{Milgrom:1983ca} and a large class of theories implementing in the action new invariants that are absent in 
the original Einstein-Hilbert action \cite{Belenchia:2016bvb}. 
On the other hand, the dark matter approach to galaxies is addressed adding new hypothetical types of matter, which interacts only gravitationally with the usual baryonic matter, inside Newton's theory, see \cite{Bertone:2018krk} for a review. \\
Usually, general relativity is not included as solution for this kind of problems, since everybody argues that the Newtonian limit of Einstein equations should be sufficient because the speeds of stars in galaxies are much smaller than the speed of light and the 
gravity is ``weak" far from the central region.
Actually, however, dealing with these considerations is far more delicate than what seems at first glance. For example, it is possible to show that steady, axially symmetric motion of dust is necessarily cylindrically symmetric according to Newtonian theory, while 
this is not necessarily true in General Relativity, \cite{Bonnor_1977}. We find it useful to report the Newtonian argument here. Consider a stationary axisymmetric system composed of dust rotating in circles around the z-axis, for which the velocity can be written as
\begin{equation}
	\Vec{v}= \omega \hat{z} \times \Vec{R},
\end{equation}
where $\Vec{R}$ is the vector position of the dust particles w.r.t. the origin of the $z$-axis, $\hat{z}$ is the versor of the $z$-axis, and $\omega= \omega(r,z)$ is the angular velocity, $r=\sqrt {\vec R^2-z^2}$. The equations of motion in the Newtonian theory are
\begin{align}
	(\Vec{v} \cdot \nabla) \vec v &=- \nabla \psi,  \label{eqV}\\
	\nabla^2 \psi&= 4 \pi \rho(r;z), \label{eqP}
\end{align}
to which we have to add the stationary form of the continuity equation
\begin{gather}
    \nabla \cdot (\rho \vec{v})=0.
\end{gather}
The $z$-component of eq.($\ref{eqV}$) gives the independence of the potential $\psi$ on the $z$ coordinate, so that differentiating $(\ref{eqP})$ we have
\begin{gather}
    \partial_z \rho=0,
\end{gather}
which means that the density gradient along $z$ must vanish! Because of these equations we have also $\omega=\omega(r)$. Then, due to the independence of the system on the $z-$coordinate we necessarily have cylindrical symmetry! This looks to be a 
strong reason for abandoning the Newtonian description of the galaxies independently from the smallness of the velocities and the weakness of the gravitational field, in favour of general theory where, instead, this result is not true and an 
axially symmetric system is not forced to become cilindrically symmetric. This has to be attributed to the higher number of degrees of freedom of general relativity with respect to gravity. \\
Up to our knowledge was in the pioneering work \cite{Cooperstock:2005qw}, by Cooperstock and Tieu (CT), and, subsequently, in \cite{Carrick:2011ac}, which General Relativity has been used for the first time to investigate the dynamics of Galaxies.  
Later on, Balasin and Grumiller introduced a new model \cite{Balasin:2006cg} (BG), eliminating the unphysical behaviour affecting some previous solutions and showing explicitly that for extended rotating sources the weak-field approximation is not Newtonian.  
This model gained relevance recently, because of \cite{Crosta:2018var} in which the authors showed perfect agreement with the Gaia DR2 catalog without the need to include dark matter.\\
Albeit the BG model is surely very interesting, it is a quite particular solution, where corotance of dust is assumed and the general degrees of freedom in axially symmetric solutions for gravitating dust are then frozen in a particular limit. By using techniques that 
can be traced to the work of Geroch \cite{Geroch:1970nt},\cite{Geroch:1972yt}, and subsequently by Hansen and Winicour  \cite{HansenWinicour1}, \cite{HansenWinicour2}, one gets that the solutions of the Einstein's equations in presence of dust can be
much more general even after imposing axial symmetry. \\
In the present work we mean to investigate the role of the additional degrees of freedom of general relativity in the galactic dynamics in the usually assumed to be a non relativistic regime. After a general discussion of it, we will apply the aforementioned general
techniques to the analysis of the curves of velocity for axisymmetric galaxies after eliminating the request of co-rotating dust, thus introducing the possibility of differential rotation. As we will show this will open even more interesting possibilities.
As shown in \cite{doi:10.1063/1.522754}, the general solution depends upon the choice of an arbitrary function $H= H(\eta)$ (see section \ref{sec:axisymmetric}). We will see that choosing $H$ corresponds to fix an equation of state for the rotation, 
relating this function to the velocity profile measured from a Minkowski observer at infinity.\\
As a particular case, we will include a new stationary solution to Einstein equation that further supports our interpretation of the non Newtonian degrees of freedom: at large distance from the center it is possible to have motion that is not supported just by the 
density of the dust but also by the momentum of the gravity. We propose that this dragging effect of the momentum of the space-time, possibly including the differential rotation, can provide an explanation to the dark matter or at least part of it. 

\section{On the energy-momentum of the gravitational field in GR} \label{sect 2}

The paradigm shift from the Newtonian gravity to the General Relativity involves the emergence of intrinsically non-Newtonian phenomena: the distortion of time flux, the gravitational lensing, the presence of black holes, the expansion of the universe, the 
existence of gravitational waves, and so on. In this work we will consider another post-Newtonian phenomenon, still not fully understood but with important manifestations, even for systems with low speeds and small energy-matter densities. \\
Let us start considering the fundamental gravitational law according to Newton%
\begin{equation}
	\Delta \Phi=4\pi G\rho
\end{equation}%
and that of Einstein%
\begin{equation}
	G_{\mu\nu}(g_{\mu\nu})=\frac{8\pi G}{c^4}T_{\mu\nu}.
\end{equation}%
From the mathematical point of view, some differences are immediately noticeable: the Newton's gravitational field $\Phi$ is scalar, while the Einstein's one is a tensor $g_{\mu\nu}$; the Newton's second order operator $\Delta$ is fully spatial, while the 
Einstein tensor $G_{\mu\nu}$ contains time derivatives; and $\Delta$ is linear, while $G_{\mu\nu}$ is not. \\
From the physical point of view, these differences means that the Einstein's gravitational field has more degrees of freedom than the Newtonian one, and that it is a dynamical field. Indeed, the gravitation of Newton is not essentially a field theory, since $\Phi$ 
is determined by the matter distribution at each instant, and the interaction can be described as an action at a distance. On the other hand, $g_{\mu\nu}$ must have all the properties of a physical, dynamical field, carrying a certain amount of energy and 
momentum independently by its source $T_{\mu\nu}$. This fact is obvious once one considers the gravitational waves, which exhibit multiple degrees of freedom with their polarizations.

A way for evaluate the energy-momentum of gravity in the GR framework was made, for example, by Landau and Lifshitz \cite{Landau:1982dva}. Defining the Landau-Lifshitz object $t^{\mu\nu}$, they found a conservation law%
\begin{equation}
	\partial_{\mu}[\sqrt{-g}(T^{\mu\nu}+t^{\mu\nu})]=0.
\end{equation}%
It seems that we can interpret $t^{\mu\nu}$ as the energy-momentum tensor carried by the gravitational field. Unfortunately, this object is not a tensor. It is possible to choose the coordinates in such a way that $t^{\mu\nu}$ vanishes on a line. Moreover, there 
exist choices of coordinates on a flat space-time for which $t^{\mu\nu}$ is not identically zero. In other words, the identification of the energy-momentum of the gravitational field is arbitrary. \\
This does not mean that it does not exist! Its relevance can be recognized from the \emph{gedankenexperiment}s of the following section.

The Newtonian description is widely believed to be a first approximation for gravitational phenomena, for which the GR gives just corrections of lower order. But, in general, this is not true for the contributes of the dynamical degrees of freedom of the field. \\
In the following sections, we will show some examples of this non negligibility.
Here we will consider three revealing examples: the non-trivial fields in the vacuum, the Carlotto-Schoen shielding, and the axisymmetric solutions. For all of these cases, non-Newtonian, non-negligible effects are shown.

\subsection{Freedoms in the vacuum}

The vacuum solutions, as well as the fields generated by a compact source, are always taken asymptotically flat. In the Newtonian theory, this admits only harmonic, asymptotically zero fields $\Phi(\underline{x},t)$, at any instant. On the other hand, the 
GR allows many shapes of vacuum solutions. \\
The black holes solutions as Schwarzschild, Kerr, Reissner-Nordstrom and Kerr-Newman are well known. Even if $T_{\mu\nu}$ is identically zero, we can say that the non-flatness of these metrics carries some field energy. An angular momentum can be 
also attributed to the rotating black holes. \\
Also the gravitational waves are vacuum solutions, analytically expressible at the first order approximation for small amplitudes. Here the dynamicity of the gravitational field is evident, and it cannot have Newtonian analogues. As a wave of any other dynamic 
field, a gravitational wave must carry an energy and a momentum. \\
There are also known non-singular, stationary vacuum solutions. This concept, firstly investigated by Wheeler, is called geon. It can be seen as a localized  gravitational wave which is held together by its own field energy. This is another evidence of the energy 
of the gravitational field, since without an energy it could not attract himself.

Even if the Landau-Lifshitz object can be locally set to zero for each one of these metrics, with the suitable choice of coordinates, the energy-momentum of these fields can be empirically detected, in some sense. Let we consider the following 
\emph{gedankenexperiment}. \\
In a vacuum universe with geon metric, we put a single test particle. If the metric had no energy, the particle could never change its velocity, because it would not have any other particle or field to interact with. However, when the particle meets the geons, its 
motion is deflected by the non-zero gravitational force inside that region. The only way to restore the energy-momentum conservation is to say that it is exchanged between the particle and the geon. Thus, the non-flatness of the space-time implies a certain 
amount of energy and momentum. \\
Another version can be done making the particle hit by a gravitational wave. For a linearized wave, the interaction with the particle is negligible. However, if the wave has a big enough amplitude to make relevant the higher terms, also its thrust to the particle 
becomes non negligible. An initially \lq\lq still\rq\rq~particle would start to \lq\lq move\rq\rq. Here, the stillness and the movement are intended with respect to the gravitational wave. This suggest that, since the metric carries a momentum, one can say that the 
matter is ``still or moving with respect to the gravity''. \\
This last conjecture can be elaborated in a third version of our \emph{gedankenexperiment}. Let us consider a photon in a Kerr space-time, traveling in the $z=0$ plane. It needs less time for a complete rotation than it would have taken in a flat space-time; this 
fact is usually called a dragging effect. Since there are no other particles to interact with, it must have gotten the increase of the angular momentum from the gravitation. Such photon is co-rotating with the Kerr metric, while an orbital motion with a different period 
would \lq\lq have a differential rotation with respect to the gravity\rq\rq.

\subsection{Shields and energy localization}
As we said, the Newtonian theory leaves quite few degrees of freedom for a field generated by a compact source: just an harmonic, asymptotically vanishing term. The GR has much more freedom. This is especially shown by the following result, due to 
Carlotto and Schoen \cite{carlotto2015localizing}.%
\begin{thm}
	Given an asymptotically flat initial data set for vacuum Einstein equations, there exist cones and asymptotically flat vacuum initial data which coincide with the original ones inside the cones and are Minkowskian outside slightly larger cones.
\end{thm}%
The meaning of this Theorem is that there exist field solutions for which any gravitational influence from the inside cone is shielded, so that the gravitational force outside is exactly zero \cite{chrusciel2016antigravity}. Notice that between the two cones there is 
again a vacuum solution, so it is not needed a material shield, as it is for the Gauss Theorem in Newtonian gravity, or for a Faraday cage. \\
The Carlotto-Schoen Theorem applies to vacuum solutions, as a black hole or a geon, but it can be easily generalized to fields generated by compact sources. Indeed, if $T_{\mu\nu}$ is identically zero outside a compact region $K$, there one has a vacuum field 
$g_{\mu\nu}$. Applying the Theorem on the complementary region $\mathbb{R}^3-K$, such that $K$ is all inside the inner cone, it is returned another vacuum field $\tilde{g}_{\mu\nu}$ which is Minkowskian outside the outer cone. Hence, the bodies in that outside 
do not feel any gravitational interaction with the source in $K$.\\
A an example, let us consider a spherical star or planet, so that $K$ is a sphere. According to Newton, a test particle at any distance would always feel the gravitational attraction, however much it descends with distance. Instead, the Einstein theory allows the 
Carlotto-Schoen solution $\tilde{g}_{\mu\nu}$, for which the test particle feels an exactly zero gravitational force in the outside region. The gravitational interaction of the star, or planet, is fully shielded.\\
This phenomenon is strongly non Newtonian, because the Newton's gravity is an action at a distance, thus the gravitation of a body cannot be shielded for another body, whatever happens between of them. The Carlotto-Shoen shielding is due to the locality of 
the relativistic theory. \\
We can interpret metrics as $\tilde{g}_{\mu\nu}$ saying that the field energy-momentum is localized inside the cones, while for the original vacuum metric $g_{\mu\nu}$ the energy-momentum is spread throughout the space. Indeed, we saw the gravitational force 
on a test particle as the exchange of energy and momentum with the field, and it cannot be done outside the cones, where the field is trivial. Even if for the given instant the field energy is localized inside the cones, the non-flat region begins to 
expand into the flat region, at the speed of light. As any other localized field carrying energy, it radially propagates from the initial region.

\subsection{Freedoms for axisymmetric and stationary rotating dust}\label{sec:axisymmetric}
An explicit class of examples can be given by a gravitational field and a distribution of dust, interacting, under the requests of stationarity and axial symmetry. The general solution has a matter density $\rho(r,z)$, a velocity field $v(r,z)$ which always 
rotates around the axis $r=0$, and a gravitational field which also can depend only on $r$ and $z$. \\
We have already remarked that Newtonian theory puts strong constrains onto the system: the density cannot depend on $z$, otherwise the gravitational attraction would make collapse the dust at the heights $z$ with bigger density, so that it would not be 
stationary.\footnote{The only exception is for the singular distribution for which all the matter and momentum are concentrated on the plane $z=0$, but it is extremely unstable.} The link between $\rho(r,z)\equiv\rho_0(r)$, $v_0(r)$ and $\Phi_0(r)$, where we can 
evaluate all of them on the plane $z=0$, is immediately given by the Gauss Theorem%
\begin{equation}
	4\pi Gr\rho=\partial_r(v^2)=\partial_r(r\Phi_r).
\end{equation}%
The degrees of freedom of this solution is given by only one single-variable function. Let now compare it to the general relativistic version.\\
A stationary, axisymmetric metric can be written in general with $g_{t\phi}$ the only non-zero off-diagonal component, and $g_{rr}=g_{zz}$,%
\begin{equation}
\label{metric}
    ds^2= -e^{2U(r,z)} \left(dt+A(r,z)d\phi\right)^2 + e^{-2U(r,z)}  \left[r^2 d\phi^2+ e^{2k(r,z)}\left(dr^2+dz^2\right)  \right]. 
\end{equation}%
As shown in \cite{stephani_kramer_maccallum_hoenselaers_herlt_2003},
it is fully fixed by the choice of a negative function $H(\eta)$, where $\eta(r,z):=rv(r,z)$, and of an axisymmetric function $\mathcal{F}(r,z)$ that satisfies the simil-harmonicity condition%
\begin{equation}
\label{harm}
	\mathcal{F}_{rr}-\frac{1}{r}\mathcal{F}_r+\mathcal{F}_{zz}=0.
\end{equation}%
The velocity field $v$, or equivalently $\eta$, can hence be found by%
\begin{equation}
\label{def v}
	\mathcal{F}=2\eta+r^2\int\frac{H'}{H}\frac{d\eta}{\eta}-\int\frac{H'}{H}\eta d\eta.
\end{equation}%
After this, the metric components take the form
\begin{align}
	&\begin{cases}
		g_{tt}&=\frac{(H-\eta\Omega)^2-r^2\Omega^2}{H} \\
		g_{t\phi}&=\frac{\eta^2-r^2}{-H}\Omega+\eta \\
		g_{\phi\phi}&=\frac{r^2-\eta^2}{-H}
	\end{cases}, \label{components} \\
	&s.t. \quad \Omega=\frac{1}{2}\int H'\frac{d\eta}{\eta} \label{eta cond};
\end{align}%
 $g_{rr}:=e^{\mu}$ is determined by the previous three equations and can be found using%
\begin{align}
    \mu_r =& \frac{1}{2r} \left[ (g_{tt})_r (g_{\phi\phi})_r-(g_{tt})_z (g_{\phi\phi})_z - ((g_{t\phi})_r )^2+((g_{t\phi})_z )^2  \right] \label{mur},\\
    \mu_z= & \frac{1}{2r} \left[ (g_{tt})_z (g_{\phi\phi})_r + (g_{tt})_r (g_{\phi\phi})_z -  2 (g_{t\phi})_r (g_{t\phi})_z  \right] \label{muz}.
\end{align}%
Finally, the matter density is%
\begin{equation}
\label{rho}
	8\pi G\rho=\frac{v^2(2-\eta l)^2-r^2l^2}{4g_{rr}}\frac{\eta_r^2+\eta_z^2}{\eta^2},
\end{equation}%
where we called $l(\eta(r,z)):=\partial_{\eta}H/H$. Along all this paragraph, the units of measure are taken s.t. $c=1$.
Since the condition on $\mathcal{F}$ leaves to it the freedom of two single-variable functions, we can immediately see that the degrees of freedom of the relativistic system are those of three single-variable functions, instead of the only one of the Newtonian
system. \\
Secondarily, the GR allows solutions without cylindrical symmetry. Instead of the Newtonian constrain $v_z=0$, another differential equation can be found for the velocity field, as we will see in the next section%
\begin{align}
\label{eq0}
    0=&l'\left(\frac{1}{v}-v\right)[r^2v_z^2+(rv_r+v)^2]+\cr
    &+l \left[\left(\frac{1}{v}-v\right)(rv_{zz}+rv_{rr}+3v_r)-\left(\frac{1}{v}+v\right)\frac{r}{v}(v_z^2+v_r^2)+\frac{2}{r}\right]+\cr
    &+2\left(v_{zz}+v_{rr}+\frac{v_r}{r}-\frac{v}{r^2}\right). 
\end{align}%
This reduces the possible shapes of $v(r,z)$, despite any choice of $l(\eta)$ can be adopted, because the single-variable function $l$ has quite less freedom than the two-variable one $v$.\\
The velocity $v(r,z)$ is defined as the velocity measured by a reference frame composed by locally non rotating observers, such as the ones used in \cite{1972ApJ}.
For a fixed $l$, (\ref{eq0}) is a non-linear second order PDE for $v$. For initial conditions $v_0(r):=v(r,0)$ and $w_0(r):=v_z(r,0)$, it defines a solution $v(r,z)$. Replacing it inside $l(rv)$ and solving for $g_{rr}$, one gets the matter distribution (\ref{rho}). Hence, 
$\rho(r,z)$ is determined from the three free functions $l(\eta)$, $v_0(r)$ and $w_0(r)$, while in the Newtonian case only $v_0(r)$ is free. \\
Again, the relativistic degrees of freedom, especially the free function $H(\eta)$, describe the space-time metric on which the matter moves. The $\Omega(r,z)$ parameter is the differential rotation of the dust with respect to the gravity, and the co-rotation is 
expressed by the choice of a constant $H$. The contribution of the field to the energy density can be seen by the terms with $l$ in the (\ref{rho}). \\
We can recognize the physical meaning of the gravitational degree of freedom $H(\eta)$ constructing suitable scalar quantities, and showing that they depend on $H$. Here we anticipate the calculation of the deformation tensor and the vorticity tensor in 
\S\ref{p2 calc}%
\begin{equation}
	\mathbf{P}= (u_{\mu;\nu}+u_{\nu;\mu})dx^\mu\otimes dx^\nu,  \quad \mathbf{W}=(u_{\mu;\nu}-u_{\nu;\mu})dx^\mu\wedge dx^\nu,
\end{equation}%
where $u=(-H)^{-1/2}(\partial_t+\Omega\partial_{\phi})$ is the four-velocity field of the dust, where $\Omega$ is given by (\ref{eta cond}). Choosing for simplicity $H(\eta):=-e^{p^2\eta^2}=-1-p^2\eta^2+o(p^2)$, we find at the lowest order
\begin{align}
	&P^2:=P_{\mu\nu}P^{\mu\nu}=2 p^4 \frac{r^2}{g_{rr}}\left(\eta_r^2+\eta_z^2\right)+o(p^4), \cr
	&W^2:=W^{\mu\nu}W_{\mu\nu}=\frac{2}{r^2g_{rr}}\left(\eta_r^2+\eta_z^2\right) (1-p^2\eta^2) +o(p^2). \label{INVA}
\end{align}%
Here it is evident the dependence on the degree of freedom $p^2$, as well as on the degrees of freedom inside $\mathcal{F}$. In particular, the co-rotating case $H\equiv-1$ shows no scalar deformation $P^2$ and we have $W^2= 2 \rho$, the density in the 
co-rotating case.

\section{Master galaxy equation}
Let us consider an observer with four-velocity $u(\tau)$ and proper clock $\tau$, so that
\begin{align}
    g(u,u)=-1.
\end{align}
The modulus of the three-velocity the observer measures for a massive particle of four-velocity $k$ is
\begin{align}
    v^2 = 1-\frac{1}{(u_{\mu} k^{\mu})^2} \label{measvel}.
\end{align}
In other words, the observer measures a Lorentz factor
\begin{align}
    \gamma= u_{\mu} k^{\mu} \label{gamma}.
\end{align}
We now assume to displace everywhere in our local neighbourhood such kind of observer so that they are at rest w.r.t. to the coordinates we considered in the previous section. 
Using (\ref{measvel}),  the velocity of the gas as measured by these observers results to be
\begin{align}
    v(r,z)= \frac{\eta(r,z)}{r} \label{vo}.
\end{align}
Therefore, it is meaningful to replace $\eta=rv$ in (\ref{harm}), then using the velocity in place of the anonymous function $\eta$.
We get
\begin{align}
\label{eq}
    0=&l'\left(\frac{1}{v}-v\right)[r^2v_z^2+(rv_r+v)^2]+\cr
    &+l \left[\left(\frac{1}{v}-v\right)(rv_{zz}+rv_{rr}+3v_r)-\left(\frac{1}{v}+v\right)\frac{r}{v}(v_z^2+v_r^2)+\frac{2}{r}\right]+\cr
    &+2\left(v_{zz}+v_{rr}+\frac{v_r}{r}-\frac{v}{r^2}\right), 
\end{align}
where  $H'= \frac{dH}{d\eta}$ and $l= \frac{H'}{H}$. See Appendix \ref{App:A} for the details. We call it {\it the Master Equation}. In general, it can be understood as a very complicate functional equation for $l$ (or for $H$) involving the velocity profile $v$ (and
its partial derivatives). Viceversa, for any given choice of $l(\eta)$, one could in principle solve it for $v(r,z)$. Starting from the form (\ref{harm}), it can also be recast in the form
\begin{align}
    \int[A(\alpha)K_1(\alpha r)+B(\alpha)I_1(\alpha r)]\cos(\alpha z)d\alpha&=\frac{\mathcal{F}(r, z)}{r}=2v+rC(rv)-\frac{1}{r}D(rv), \cr
    s.t. \quad \eta C'(\eta)&=l(\eta)=\frac{1}{\eta}D'(\eta),
\end{align}
where $A, B$ are arbitrary functions. For instance, if we take the simple choice $H\propto\eta^K$, with constant $K$, the velocity profile $v$ becomes the solution of the second degree algebraic equation
\begin{equation*}
    (2-K)v^2-\frac{\mathcal{F}}{r}v-K=0,
\end{equation*}
which, in particular, for $K=2$ returns $v=\frac{2}{\mathcal{F}}$. Another possible interesting case is the choice $H=-e^{p^2\eta^2}=-(1+p^2\eta^2)+o(p^2)$, (see \S\ref{p2 calc}) for which we have to solve the third degree algebraic equation
\begin{equation*}
    \frac{2}{3}p^2r^2v^3-2(1+p^2r^2)v+\frac{\mathcal{F}}{r}=o(p^2).
\end{equation*}
Finally, a choice of the form $H=- e^{l_0\eta}$ gives the non-algebraic equation
\begin{equation*}
    \frac{1}{2}v^2-2v-l_0r\ln v+\frac{\mathcal{F}}{r}=0.
\end{equation*} \\
A particular solution of the Master Equation, applicable to finite regions of $r$, is showed in Appendix \ref{v constant}. The geodesic equations are solved by particles going trough $r=~\text{constant}$ orbits, with constant speed $v$. It admits a vacuum subclass 
of solutions. This is another very interesting example of a non-trivial vacuum solution, that confirms the role of the gravitational field we described in the previous Section. \\
This freedom in choosing $H$, or $A$ and $B$, does not allow to reproduce all possible profiles for $v(r,z)$, so that the master equation restricts the physically admissible profiles compatible with GR and axial symmetry. \\
The co-rotating case $\Omega=0$ is obtained as the limit case of constant $H$, like in the BG model, for which we get
\begin{align}
v_{zz}+v_{rr} - \frac{v}{r^2} + \frac{v_r}{r}=0.
\end{align}
This equation is equivalent to the Laplace equation in a flat space, using cylindrical coordinates $(r,z,\phi)$ and posing
\begin{align}
\tilde v(r,z,\phi)= v(r,z) e^{i \phi}.
\end{align}
Actually, the co-rotating solution in our framework cannot be directly obtained just posing $H=$constant, but, instead, we can choose $H=-1+p^2 \eta^2$ and after computing the quantities of interest, we can take the limit $p\rightarrow 0$, so the four-vector 
for the dust becomes $u= \partial_t$.\\
In order to distinguish between the co-rotance and non co-rotance, let us compare the precession of the intrinsically non-rotating observers in both models with respect to the gyroscopes they bring with themselves. An intrinsically non rotating 
orthonormal frame is
\begin{gather}
    e^0= \frac{r}{\sqrt{g_{\phi\phi}}} dt, \qquad  e^1= \frac{(d\phi-\chi dt)}{\sqrt{g_{\phi\phi}}},\qquad e^2= e^{\mu/2}\, dr,\qquad e^3= e^{\mu/2}\, dz,
\end{gather}
where, for simplicity, we defined
\begin{gather}
\chi \equiv - \frac{g_{t\phi}}{g_{\phi\phi}}=  \frac{H \eta}{(r^2-\eta^2)}+\Omega.    
\end{gather}
Applying (\ref{measvel}) to $e^0$ leads to (\ref{vo}). The relevant elements of the dual basis are found to be
\begin{gather}
   e^0 \rightarrow X\propto \left( \partial_t+ \chi \partial_\phi \right),  \\
   e^1 \rightarrow Y \propto \partial_\phi,
\end{gather}
showing that the observers rotate with angular velocity equal to $\chi$ in these coordinates.
The connection 1-forms are 
\begin{align}
    \omega^0_{\,a}&= - \left(\partial_a \log\frac{r}{\sqrt{g_{\phi\phi}}} \right) \frac{e^0}{\sqrt{g_{rr}}}+ \frac{1}{2} \frac{\partial_a \chi}{\sqrt{g_{rr}}\,r} e^1, \\
    \omega^0_{\,1}&= \frac{1}{2} \frac{1}{\sqrt{g_{rr}}\,r} \left(\partial_r \chi\, e^2+\partial_z \chi \,e^3 \right), \\
    \omega^2_{\,3}&= \left(-\left(\partial_z \log \sqrt{g_{rr}} \right) e^2+ \left(\partial_r \log \sqrt{g_{rr}} \right) e^3 \right), \\
    \omega^a_{\,1}&= \left(\partial_a \log\frac{1}{\sqrt{g_{\phi\phi}}} \right) \frac{e^1}{\sqrt{g_{rr}}}- \frac{1}{2} \frac{\partial_a \chi}{\sqrt{g_{rr}}\,r} e^0,
\end{align}
where $a=(2,3)$. From here one can see that this reference frame is intrinsically non-rotating, although it is non inertial since the three-acceleration it experiences is 
\begin{gather}
    \vec{a} \propto  \partial_\alpha \log(\frac{r}{\sqrt{g_{\phi\phi}}}), \qquad \alpha=r,z,
\end{gather}
which is the force required by its thrusters to keep this observer in its orbit. The corresponding gyroscopes precess relatively to the orthonormal frame with angular velocity
\begin{gather}
   \omega \propto \partial_\alpha \chi, \qquad \alpha=r,z.
\end{gather}
The precession of the gyroscopes respect to the intrinsically non-rotating observers in the non-co-rotating and co-rotating cases can be now compared as
\begin{gather}
    \Delta \omega=\omega_{nCor}-\omega_{Cor} \propto \frac{\eta}{r^2-\eta^2} \partial_\alpha H+ \partial_\alpha \Omega+ \Delta H \, \partial_\alpha \left(\frac{\eta}{r^2-\eta^2}\right),
\end{gather}
where $\Delta H= H_{nCor}- H_{Cor}$. 
The precession measure can distinguish between the two situations.

\section{Behaviour far from the center} \label{BFFC}

Let us now analyse the density in eq.$(\ref{rho})$, where we remember that $\eta=v r$ and $\kappa= 8\pi G$.
We want to study the behaviour of this system far from the bulge, in the external part, where the velocity is observed to be almost constant and the dark matter effects look more relevant. 
The full solution of eq. $(\ref{eq})$ is very hard to achieve and should tackled numerically. Nevertheless, we can get some important insight from general considerations. As already stressed, the third line of eq. $(\ref{eq})$ describes the galaxy in the co-rotating 
approximation, so, to understand the physics after relaxing this condition, we can start considering $l$ approximately constant, such that it is comparable to a characteristic velocity/length of the system. Therefore, compatibly with $\rho >0$, we choose
\begin{gather}
    l= \frac{a v_{c}}{R_G}, 
\end{gather}
where  $v_c$ is order $10^{-4}$ and $R_G$ is comparable to the radius of the galaxy, or even larger if we need to describe further regions; in that case we have to choose $a$ appropriately. These simple choices are enough to reach interesting conclusions. 
We have
\begin{gather}
\label{density}
    4 g_{rr} \kappa \rho= \frac{1}{v^2 r^2} \left[ v^2 \left(4-a^2 \left(\frac{r}{R_G}\right)^2 \left(\frac{v_c}{v}\right)^2 \right) ((\eta_r)^2+(\eta_z)^2) \right],
\end{gather}
where we neglected higher order terms in $v^2$.
Notice that $a=0$ gives the co-rotating case, while $a \neq 0$ gives non co-rotance. With constant $l$, at the first non-zero order in $v$ (\ref{eq}) becomes
\begin{gather}
     \left( 2+a \frac{r}{R_G} \frac{v_c}{v}\right) (v_{zz}+v_{rr})+ \left(\frac{3a}{R_G}  \frac{v_c}{v}+\frac{2}{r} \right)v_{r}- a  \frac{v_c}{v^2} \frac{r}{R_G}((v_z)^2+(v_r)^2)+ \frac{2a}{R_G r} v_c -\frac{2}{r^2} v=0. \label{eq1}
\end{gather}
In the limit $r \ll R_G$, in the inner part of the galaxy, we get 
\begin{gather}
    v_{zz}+v_{rr}+ \frac{v_r}{r}- \frac{v}{r^2}=0,
\end{gather}
as in the BG case. Therefore, the discrepancy with the co-rotating case arise as we approach the \lq\lq edges\rq\rq~of the galaxy. Indeed, for $r \approx R_G$ we get
\begin{gather}
    \left(2+a  \frac{v_c}{v} \right) (v_{zz}+v_{rr})+\left(3a  \frac{v_c}{v} +2\right)\frac{v_r}{r}-a\frac{v_c}{v^2}((v_z)^2+(v_r)^2)+ 2\left(a  \frac{v_c}{v}-1\right) \frac{v}{r^2}=0.
\end{gather}
Remarkably this approximated equation explicitly allows for the case $v\equiv av_c$, which gives us constant velocity. Far from the center of the galaxy the observed dust velocity is nearly constant. We now assume a slight deviation from co-rotation, and, in 
particular, that $l$ and $v$ are small and constant. To be explicit we have
\begin{gather}
    H=- e^{lvr} \approx -(1+lvr), \qquad  H'= \frac{dH}{d\eta}= - l e^{lvr}\approx -l, \qquad H''=0.
\end{gather}
The angular velocity $\Omega$ is obtained as in eq.$(\ref{eta cond})$
\begin{gather}
    \Omega=- \frac{l}{2} \int \frac{e^{v l r}}{r} \approx  - \frac{l}{2} \log\left(\frac{r}{r_0}\right) ,
\end{gather}
where we are neglecting squared velocities, and $r_0$ is an integration constant. The $g_{tt}$, $g_{t\phi}$, $g_{\phi\phi}$ elements of the metric, in eq.(2.2)-(2.4), are
\begin{align}
    g_{tt}&\approx -1, \cr
    g_{t\phi}&\approx \frac{l}{2} \log(r/r_0) r^2 + v r, \cr
    g_{\phi\phi}&\approx r^2. \label{gpp}
\end{align}
From (\ref{mur})-(\ref{muz}) we find the last element of the metric
\begin{gather}
    \mu_r \approx- \frac{1}{2r} \left[ \frac{a}{2} v_c \frac{r}{R_G} \left(1+ 2 \log\left(\frac{r}{r_0}\right)\right)+v  \right]^2 ,
\end{gather}
which gives $\mu$ approximately constant. As in the co-rotating case, we will call this constant $e^{\mu}=\zeta^{-1}$. Up to order $v$, the metric and the four-velocity of the dust read\footnote{Notice that in the given approximation $N^2$ is negligible compared to
$r^2$}
\begin{align}
    ds^2 &=- \left(dt- N d\phi \right)^2 + r^2 d\phi^2+ \zeta \left(dr^2 +dz^2 \right),  \\
    N&=\frac{a}{2} v_c \log(r/r_0) \frac{r^2}{R_G} + v  r,\\
   u&= \left( \partial_t - \frac{a}{2} \frac{v_c}{R_G} \log\left(\frac{r}{r_0}\right) \partial_\phi\right).
\end{align}
Then, the density reduces to
\begin{gather}
    \kappa \rho= \zeta \frac{\left(4-a^2 (r/R_G)^2 (v_c/v)^2 \right)}{4} \frac{((\eta_r)^2+(\eta_z)^2)}{r^2}.
\end{gather}
We can compare co-rotating and non co-rotating densities by assuming the same velocity profiles:
\begin{gather}
    \frac{\rho_{nCor}}{\rho_{Cor}}= \frac{\left(4-a^2 (r/R_G)^2 (v_c/v)^2 \right)}{4} \approx 1-\frac{a^2}{4} \left(\frac{r}{R_G}\right)^2 \left(\frac{v_c}{v}\right)^2.\label{ncorvscor}
\end{gather}
In \cite{Crosta:2018var} it has been shown that the co-rotating model of the Milky way already fits the Gaia's data without need of any dark matter (at the present value of precision). This means that for our galaxy $a$ is very small. However, for other galaxies 
non-corotance may be more important, leading to a further reduction of the needed density, as in (\ref{ncorvscor}), then showing apparently a larger amount of dark matter.
Therefore, it looks exactly as the gravitational field itself is pulling the dust, carrying its own inertia. However, to push forward such interpretation it would be interesting to understand more deeply the energy momentum pseudo tensor of gravity, for example 
the Landau-Lifshitz affine-tensor, \cite{Landau:1982dva}. 

\section{Almost constant velocity and small co-rotation}
\label{p2 calc}

Here we consider some acceptability conditions for the metric (\ref{metric}). 
Of course, the solution we are considering is constrained into the region where matter density is non negative. Remembering (\ref{density}), we can write
\begin{equation}
    4\kappa e^{2k}\rho=(-g_{tt})\frac{v^2(2-\eta l)^2-r^2l^2}{\eta^2}(\eta_r^2+\eta_z^2),
\end{equation}
and that $-g_{tt}>0$ in the halo, $\rho\geq0$ is equivalent to
\begin{equation}
\label{dens hor}
    rl\leq v|2-\eta l|.
\end{equation}
This expresses a possible restriction on the shapes of $v$ or $l$. E.g. for the choice $H:=-e^{p^2\eta^2} \Leftrightarrow l=2p^2\eta$, we get
\begin{equation*}
    r\leq\frac{1}{p\sqrt{1+v^2}}\leq\frac{1}{p}:=R_M,
\end{equation*}
but it can be written also as
\begin{equation*}
    v(r, z)\leq\frac{\sqrt{1-p^2r^2}}{pr}=\sqrt{\frac{R_M^2}{r^2}-1}.
\end{equation*}
This constraints the values of $p$ to $p<1/R_G$.
Therefore, we can approximate
\begin{align}
    H= - \left( 1+p^2 \eta^2 \right)+o(p^2\eta^2), \label{h}
\end{align} 
since $p\eta\approx pR_G v<v$.
For convenience, however, we start with the exact expression $- \left( 1+p^2 \eta^2 \right)$ for $H$, and after we expand in $p^2$. From (\ref{def v}) we get
\begin{align}
    \eta(r,z) = \frac{1}{p} \tan\left(  \frac{p \,\mathcal{F}(r,z)}{2(1+p^2 r^2)}\right)\label{2}.
\end{align}
The corresponding solution is
\begin{align}
  ds^2=&-\frac{1 -\eta^2 p^4 r^2 }{1+p^2 \eta^2}  \left[ dt- \eta \frac{(1+p^2 r^2) }{ 1 -\eta^2 p^4 r^2  }  d\phi \right]^2 + \frac{ 1+p^2 \eta^2}{1 -\eta^2 p^4 r^2} \left[ r^2 d\phi^2 + e^{2k}\left( dr^2 +dz^2 \right)\right] \label{Metric}, \\
  \Omega(r,z) =& \Omega_0 - p^2 \eta(r,z) \label{Omega},\\
    \kappa e^{2k} \rho =& \frac{1}{r^2}\frac{(1-p^4 r^4) \left( 1+p^2 \eta^2\right)}{(1- p^4 \eta^2 r^2)(1+p^2 r^2)^2}\left[\mathcal{F}^2_z+ \left(\mathcal{F}_r - 2 \frac{p^2 r}{1+p^2r^2} \mathcal{F}\right)^2\right] \label{Density},
\end{align}
and
\begin{align}
 e^{2k}=\zeta \frac {1-p^4 r^2 \eta^2}{1+p^2\eta^2} \exp \left[\int \frac {(1+p^2r^2)^2 \left(\eta_z^2-\eta_r^2 \right)}{2r(1+p^2\eta^2)^2} dr \right].
\end{align}
Expanding in $p^2$, (\ref{2}) and (\ref{Omega})  simplify to 
\begin{align}
    \eta=& \mathcal{F} \left[1+p^2\left( \frac{ \mathcal{F}^2}{3}- r^2 \right)\right]+o(p^2), \\ \Omega=& -p^2 \mathcal{F} +o(p^2),
\end{align}
where we absorbed a factor $\frac{1}{2}$ in $\mathcal{F}$.
The metric (\ref{Metric}), the density (\ref{Density}) and the four-velocity of the dust read
\begin{align}
    ds^2=& -\left(1-p^2 \mathcal{F}^2\right) \left[ dt- \mathcal{F} \left(1+ \frac{p^2}{3} \mathcal{F}^2\right)d\phi\right]^2 + \nonumber \\ &+\frac{1}{\left(1-p^2 \mathcal{F}^2\right)} \left[ e^{2k}\left(dr^2+dz^2\right)+ r^2 d\phi^2 \right] + o(p^2),\\
e^{2k} \rho= &\frac{1}{r^2}\left(1-2p^2 r^2\right) \left(1+p^2 \mathcal{F}^2\right)\left[ \mathcal{F}_z^2+ \left( \mathcal{F}_r -2p^2 r \mathcal{F}\right)^2\right] +o(p^2), \\   
    u=& \left(1-\frac{p^2}{2}  \mathcal{F}^2 \right) \left( \partial_t - p^2 \mathcal{F} \partial_{\phi}\right) +o(p^2). \label{u}
\end{align}
To obtain this limit we implicitly required
\begin{align}
    \frac{1}{\mathcal{F}(r,z)}> p \label{C3},
\end{align}
since $\mathcal F \geq \eta$. 
From (\ref{measvel}) we find the velocity
\begin{align}
    v= \frac{\mathcal{F}}{r} \left[1+p^2 \left(\frac{\mathcal{F}^2}{3}-r^2\right)\right] \label{vop2},
\end{align}
which gives the correction with respect to the co-rotating case. Notice that one could start from the different choice $H=-e^{-p^2\eta^2}$. The corresponding solution can be easily get simply replacing $p^2$ with $-p^2$ everywhere, so we replace $p^2$
with $\epsilon p^2$, $\epsilon$ a sign. 
In this approximation $\mathcal F$ is independent from $p$. Since for $p=0$ we want to reproduce the BG solution, we have to take for $\mathcal F$ the same expression as in \cite{Balasin:2006cg}, (their $N(r,z)$ in (25)). In particular,
in the almost constant velocity ($V_0$ in the notation of \cite{Balasin:2006cg}) region we get
\begin{align}
 v\simeq V_0 \left( 1-\epsilon p^2 r^2 \left(1-\frac {V_0^2}3\right) \right).
\end{align}

\subsection{Deformation and Whirling tensor up to $p^2$}
In order to compare the co-rotating solutions with the non co-rotating ones it is interesting to introduce the deformation tensor $ \mathbf{P}$ 
\begin{align}
    & \mathbf{P}(u) \equiv \mathcal{L}_u(g)= (u_{\mu;\nu} + u_{\nu;\mu}) dx^\mu \otimes dx^\nu,  \label{deformation}
\end{align}
where $\mathcal{L}$ is the Lie derivative. Up to order $p^2$ it reads
\begin{align}
    \mathbf{P}(u)=& -p^2r^2 \left( \mathcal{F}_r\, dr\odot d\phi +\mathcal{F}_z \,dz\odot d\phi \right)= \nonumber \\=& -p^2 r^2 d\mathcal{F} \odot d\phi.
\end{align}
As expected, the behaviour in the co-rotating case $p \rightarrow 0$ is that of a ``rigid fluid.'' Notice that the deformation is only ``spatial'' in the proper reference frame of the dust.
This tensor gives the invariant 
\begin{align}
   \mathbf{P}^2 \equiv \mathbf{P}_{\mu\nu}\mathbf{P}^{\mu\nu}= 2p^4 r^2 \left( \mathcal{F}_r^2+\mathcal{F}_z^2\right).
\end{align}
This invariant is exactly zero in the co-rotating case.
It may be of interest also the vorticity $\mathbf W$, defined as
\begin{align}
    & \mathbf{W}(u) :=  du= (u_{\mu;\nu} - u_{\nu;\mu}) dx^\mu \wedge dx^\nu, \label{Whirl}
\end{align}
which has components
\begin{align*}
\mathbf{W}_{tr}=& -p^2 \mathcal{F}\mathcal{F}_r, \quad \mathbf{W}_{tz}=-p^2 \mathcal{F}\mathcal{F}_z, \\
\mathbf{W}_{r\phi}=& -\mathcal{F}_r \left[1-p^2 \left(r^2+ \frac{\mathcal{F}^2}{2}\right) \right] +2p^2 r \mathcal{F}, \\
\mathbf{W}_{z\phi}=& -\mathcal{F}_z \left[1-p^2 \left(r^2+ \frac{\mathcal{F}^2}{2}\right) \right].
\end{align*}
It gives the invariant
\begin{equation}
  \mathbf{W}^2 := \mathbf{W}^{\mu\nu}\mathbf{W}_{\mu\nu} =\frac{2 e^{-2k}}{r^2 }\left[ \left(1-2p^2 r^2-5p^2 \mathcal{F}^2 \right)\left(\mathcal{F}_r^2+\mathcal{F}_z^2\right)-4 p^2 r \mathcal{F}\mathcal{F}_r\right].
\end{equation}

\section{Final remarks}
Since \cite{Crosta:2018var} shows that the Milky Way can be well described by the Balasin-Grumiller model, we can conclude that our galaxy is almost co-rotating with the gravity. This is someway intuitive, because it is a big and old galaxy, so that the matter 
and the field had enough time to exchange angular momentum, possibly reaching the same rotation. Such a mechanism would include a breaking of stationarity and would deserve further study, which we mean to tackle in a following work. If true,
it would also suggest an higher probability to find more dark component in younger galaxies, a fact that could be investigated.
It should be checked also the role of the average rotation for elliptic galaxies, what requires a statistical study. \\
More work is in order, which could possibly allows us to better understand the dynamics introduced by the momentum and inertia of the gravitational field, providing new explanations for some unsatisfactory models.

\section*{acknowledgments}
We want to acknowledge Vittorio Gorini for drawing our attention to reference \cite{Crosta:2018var}. We also thank Maria Teresa Crosta and Alexander Kamenshchik.

\newpage
\appendix

\section{Deriving the master equation} \label{App:A}
In this section we will deduce the master equation.
Starting from eq.(\ref{harm}) we can write
\begin{align}
    \sum_a   \partial_a  \left[ 1/r \left[(r^2 \beta)_a + 2 \eta_a -  \frac{\eta}{H} H_a \right] \right]=\partial_a(\epsilon^{ab}\gamma_b)=\gamma_{rz}-\gamma_{zr}=0.
\end{align}
In order to write down the left hand side of this expression explicitly, it is convenient to we put $\eta(r,z)= v(r,z) \, r$ as in (\ref{vo}), so that we get
\begin{align*}
    0=&\partial_z\left[\frac{1}{r}\left[(r^2\beta)_z+2\eta_z-\frac{\eta}{H}H_z\right]\right]+\partial_r\left[\frac{1}{r}\left[(r^2\beta)_r+2\eta_r-\frac{\eta}{H}H_r\right]\right] \frac{1}{r}\partial_z\left[r^2\beta_z+2\eta_z-\frac{\eta}{H}H_z\right]\cr
    &+\frac{1}{r}\partial_r\left[2r\beta+r^2\beta_r+2\eta_r-\frac{\eta}{H}H_r\right]-\frac{1}{r^2}\left[2r\beta+r^2\beta_r+2\eta_r-\frac{\eta}{H}H_r\right]\cr
    =&\frac{1}{H}\left(\frac{\eta}{r}-\frac{r}{\eta}\right)\left[\frac{H_z^2+H_r^2}{H}-(H_{zz}+H_{rr})\right]-\left(\frac{\eta}{r}+\frac{r}{\eta}\right)\frac{H_z\eta_z+H_r\eta_r}{H\eta}+\cr
    &+\left(\frac{\eta}{r}+3\frac{r}{\eta}\right)\frac{H_r}{rH}+\frac{2}{r^2}\left[r(\eta_{zz}+\eta_{rr})-\eta_r\right].
\end{align*}
Remembering that $H(r,z)=H(\eta(r,z))$, we can replace $H_a=H'\eta_a$, $H_{aa}=H''\eta_a^2+H'\eta_{aa}$, where a assumes the values $r, z$, in order to get
\begin{align*}
    0=&\frac{1}{H}\left(\frac{\eta}{r}-\frac{r}{\eta}\right)\left[\frac{(H'\eta_z)^2+(H'\eta_r)^2}{H}-(H''\eta_z^2+H'\eta_{zz})-(H''\eta_r^2+H'\eta_{rr})\right]\cr
    &-\left(\frac{\eta}{r}+\frac{r}{\eta}\right)\frac{H'\eta_z\eta_z+H'\eta_r\eta_r}{H\eta}+\left(\frac{\eta}{r}+3\frac{r}{\eta}\right)\frac{H'\eta_r}{rH}+\frac{2}{r^2}\left[r(\eta_{zz}+\eta_{rr})-\eta_r\right]\cr
    =&(\ln |H|)''\left(\frac{r}{\eta}-\frac{\eta}{r}\right)(\eta_z^2+\eta_r^2)+(\ln |H|)'\left[\left(\frac{r}{\eta}-\frac{\eta}{r}\right)(\eta_{zz}+\eta_{rr}) \right. \cr
    &-\left. \left(\frac{r}{\eta}+\frac{\eta}{r}\right)\frac{\eta_z^2+\eta_r^2}{\eta}+\left(3\frac{r}{\eta}+\frac{\eta}{r}\right)\frac{\eta_r}{r}\right]\cr
    &+\frac{2}{r^2}\left[r(\eta_{zz}+\eta_{rr})-\eta_r\right].
\end{align*}%
We replace $\eta(r, z)=rv(r, z)$, for which $\eta_r=rv_r+v$, $\eta_{rr}=rv_{rr}+2v_r$. We have%
\begin{align*}
    0=&(\ln |H|)''(\frac{1}{v}-v)[ (v+r v_r)^2 +r^2 v_z^2]\cr
    &+(\ln |H|)'\left[\frac{r}{v}\left(v_{rr}+v_{zz}-\frac{v_r^2+v_z^2}{v}\right)-rv(v_{zz}+v_{rr}) - r(v_z^2+v_r^2)+3\frac{v_r}{v}-3v_rv+\frac{2}{r}\right]\cr
    &+\frac{2}{r}[2v_r+r(v_{zz}+v_{rr})]-\frac{2}{r^2}(v+rv_r).
\end{align*}

\section{Constant velocity solution} \label{v constant}

We look for solutions with constant speed for the dust. For constant $v$, we get
\begin{align}
    \frac{(1-v^2)}{2} \frac{d^2l}{d\eta^2}+ \frac{1}{\eta} \frac{dl}{d\eta}- \frac{v^2}{\eta^2}=0.
\end{align}
This is an Euler equation with constant source, whose solution is 
\begin{align}
    H(r,z)= -B \, r^\frac{2v^2}{1+v^2}  \exp(-A \,{r^{1-2\gamma^2}}),
\end{align}
where $\gamma$ is the Lorentz factor for $v$ and $A,B$ are positive constants.  The corresponding metric is
\begin{align}
\label{metr v}
g_{tt}=& -\frac{B}{4}  \frac{1-v^2}{v^2} r^{\frac{2 v^2}{v^2+1}}
   e^{-A r^{\frac{v^2+1}{1-v^2}}} \times \nonumber \\
   &
   \left(2 
   e^{A r^{\frac{v^2+1}{1-v^2}}}
   E+\frac{\left(1-v^2\right)^2
   E^2}
   {\left(1+v^2\right)^2}+ e^{2 A
   r^{\frac{v^2+1}{1-v^2}}}\right) \nonumber \\
   g_{t\phi}=& \frac{r}{2
   \left(v^3+v\right)}   \left[\left(1-v^2\right)^2
   e^{A r^{\frac{v^2+1}{1-v^2}}}
   E+\left(v^2+1\right)^2\right]\nonumber \\
   g_{\phi\phi}=& -\frac{\left(1-v^2\right)}{B} r^{\frac{2}{v^2+1}} e^{-A
   r^{\frac{v^2+1}{1-v^2}}}
\end{align}
where $E$ is the exponential integral
\begin{align}
   E= E_{\frac{4v^2}{\left(v^2+1\right)^2}}\left(Ar^{-\frac{v^2+1}{1-v^2}}\right) \quad s.t. \quad E_{n}(z)=\int_1^\infty e^{-z t}/t^n dt.
\end{align}

\bibliographystyle{ieeetr}
\bibliography{refs}

\end{document}